\documentclass[10pt]{iopart}

\usepackage{cite}
\usepackage[english]{babel}
\usepackage[utf8]{inputenc} 
\usepackage[table]{xcolor} 
\usepackage[%
  colorlinks=true,
  urlcolor=blue,
  linkcolor=blue,
  citecolor=blue
]{hyperref}
\usepackage{algorithm} 
\usepackage{algpseudocode} 
\usepackage{bbold} 
\usepackage{xcolor} 
\usepackage{graphics} 
\usepackage{amssymb}
\usepackage{bm}
\usepackage{bbm}
\usepackage{tabularx, booktabs}
\usepackage{xspace}
\usepackage{listings}
\usepackage{lipsum}
\usepackage{chemformula} 
\usepackage{braket}
\usepackage[lofdepth]{subfig}
\newcolumntype{Y}{>{\centering\arraybackslash}X}
\definecolor{dgreen}{rgb}{0,.5,0}
\definecolor{dblue}{rgb}{0,0,.5}
\definecolor{dred}{rgb}{0.5,0,.5}
\usepackage{soul}
\usepackage{diagbox}
\usepackage{setspace}

\usepackage{lmodern}
\usepackage{glossaries}
\usepackage{amsmath, amssymb}
\usepackage{mathtools}
 \addtocontents{toc}{\protect\thispagestyle{empty}}
\usepackage{mathabx}
\usepackage{listings}
\usepackage{amsmath}
\usepackage{bm}
\usepackage{parskip}
\usepackage{physics}
\usepackage{multirow}
\usepackage{cancel}
\usepackage[table,dvipsnames]{xcolor}
\usepackage{chemscheme}
\usepackage{graphicx}
\usepackage{bbold}
\usepackage[utf8]{inputenc}
\usepackage[english]{babel}
\usepackage{booktabs,tabularx}
\usepackage{appendix}
\usepackage[T1]{fontenc}
\usepackage[normalem]{ulem}

\usepackage{listingsutf8}
\usepackage{wrapfig}
\usepackage[lofdepth]{subfig}
\usepackage{verbatim}
\usepackage{multicol}
\usepackage{float}
\usepackage{stmaryrd}
\usepackage{dcolumn}
\usepackage{bm}
\usepackage{glossaries}
\usepackage{acronym} 
\usepackage{blindtext} 
\usepackage{epigraph} 
\usepackage{subcaption} 

\begin{document}

\title{A State-Specific Iterative Decoupling Scheme Based on Perturbation Theory for Low-Energy Electronic States
}

\author{Oussama Bindech*}
\ead{obindech@unistra.fr}
\address{Laboratoire de Chimie Quantique, Institut de Chimie,
CNRS/Université de Strasbourg, 4 rue Blaise Pascal, 67000 Strasbourg, France}

\author{Saad Yalouz}
\address{Laboratoire de Chimie Quantique, Institut de Chimie,
CNRS/Université de Strasbourg, 4 rue Blaise Pascal, 67000 Strasbourg, France}

\author{Vincent Robert*} 
\ead{vrobert@unistra.fr}
\address{Laboratoire de Chimie Quantique, Institut de Chimie,
CNRS/Université de Strasbourg, 4 rue Blaise Pascal, 67000 Strasbourg, France}




\begin{abstract}  
In this work, we introduce a selective and scalable extension of the multi-step Rayleigh–Schrödinger and Brillouin–Wigner (RSBW) perturbative scheme (see Ref. ~\cite{bindech2024combining}) 
, designed to efficiently access the low-energy spectrum of molecular systems. The method proceeds by combining successive effective Hamiltonian diagonalizations inspired by second-order Rayleigh–Schrödinger perturbation theory, with a Brillouin–Wigner correction applied to individually optimized states using an updated partitioning of the Hamiltonian.
At each step, a zeroth-order state is identified and progressively decoupled from the remaining higher-lying states, thereby enabling a well-conditioned Brillouin–Wigner expansion for the energy correction. In contrast to previous approaches,  
the method selectively targets a small number of low-lying states, significantly reducing the numerical overhead while maintaining spectroscopic accuracy. The robustness of the method is demonstrated on the LiH and H$_4$ molecules, where accurate excitation energies are obtained for the lowest singlet states using compact model spaces, confirming its potential for realistic applications.
\end{abstract}

 \maketitle
 \ioptwocol


\section{Introduction} 
Accurately predicting low-lying electronic spectra remains a central challenge in quantum chemistry. 
Traditional approaches, rooted in the Configuration Interaction (CI) framework, typically rely on diagonalizing the electronic Hamiltonian (or parts of it) 
to obtain electronic energies and wavefunctions. 
Without truncation, complete diagonalization  yields the so-called Full Configuration Interaction (FCI) expansion, which in principle provides exact solutions within a given basis and thus represents the most accurate approach. 
However, the exponential scaling of FCI with the system size severely restricts its applicability to small systems, rendering it impractical for realistic molecular studies (although recent works have shown that with extremely large computational resources new records can still be achieved in this direction, see Refs.~\cite{shayit2025breaking,gao2024distributed}).
To mitigate this limitation, the Complete-Active-Space (CAS) approximation reduces the cost 
by restricting the expansion to determinants generated within a selected set of active orbitals.
This framework underlies widely used approaches such as the CAS Self-Consistent-Field method (and extensions)~\cite{siegbahn1980comparison,roos1980complete,roos1982simple,malmqvist1990restricted,malmqvist2002restricted} which have become reference tools of modern computational quantum chemistry.
However, the precision of CAS-based methods strongly depends on the choice of an appropriate active space for the calculation~\cite{veryazov2011select}.
The active space determination is still regarded as an open issue in the electronic structure community and represents the main conceptual limitation of CAS-based methods, even if  recent works have focused on developing strategies for automated active space selection\cite{king2021ranked,stein2016automated,sayfutyarova2017automated,stein2019autocas} and systematic active space expansion~\cite{smith2017cheap,levine2020casscf,vogiatzis2017pushing} (note also the existence of complementary works focusing on quantum information analysis of wavefunctions and orbitals~\cite{ding2023quantum,ding2020concept,ding2025entanglement,ding2020correlation}).

As an alternative to brute-force CI expansions, Perturbation Theory (PT) formalisms have proven widely useful for addressing the electronic structure problem.
PT can refine an initial calculation into a more accurate treatment, either based on a mean-field reference, leading to the well-known Møller–Plesset approach~\cite{moller1934note}, or on an active-space reference, giving rise to CASPT2~\cite{roos1982simple,andersson1990second} and NEVPT2~\cite{angeli2001introduction,angeli2001n,angeli2002n} methods.
PT has also been extensively applied in approaches such as CI using perturbative selection done iteratively, i.e., the so-called CIPSI approach~\cite{huron1973iterative,damour2024selected,burton2024rationale,eriksen2020shape}, originally proposed to mitigate the computational cost of FCI by guiding selected CI expansions with a perturbative criterion. 
Building on this idea, the CIPSI scheme was later complemented by the intermediate Hamiltonians approach~\cite{malrieu1985intermediate,garniron2018selected}. These examples illustrate that, beyond serving as a practical tool to refine existing calculations, PT formalisms can also provide a powerful  framework for progressively constructing more sophisticated wavefunctions.

Motivated by the flexibility of the PT formalism, in this work we discuss an alternative approach 
grounded in the so-called Quasi-Degenerate Perturbation Theory (QDPT)~\cite{lindgren2012atomic, van1929sigma, kemble1937fundamental}.
 Following QDPT, one traditionally splits 
the CI space into the model $P$ space (spanned by 
the $ | \psi_{\alpha }  \rangle$ configurations with energy $E_\alpha$) and the orthogonal $Q$ space
(spanned by the $ | \psi_{\beta }  \rangle$ configurations with energy $E_\beta$)
and builds the associate effective Hamiltonian operator.
Within the Bloch-Rayleigh-Schrodinger flavor of QDPT~\cite{lindgren2012atomic} (which will be central in this work), the resulting form of the second-order effective Hamiltonian within the $P$ space reads as follows:
\begin{equation}
\begin{aligned}
\langle \psi_\alpha |\hat{H}^{\rm eff}_{2}| \psi_{\alpha '}  \rangle 
&= \delta_{\alpha\alpha '} E_{\alpha '} + \langle \psi_\alpha  |\hat{W} | \psi_{\alpha '}  \rangle \\
& + \underset{\beta}{\sum} \frac{  \langle \psi_\alpha |\hat{W} | \psi_{\beta}  \rangle  \langle \psi_\beta  |\hat{W} | \psi_{\alpha '}  \rangle  }{E_{\alpha'} - E_\beta},
\end{aligned} 
    \label{eq:effective-hamiltonian}
\end{equation}
where $\hat{W}$ is the pertubation operator.

Starting from such a space decomposition, recently a two-step scheme has been proposed
to combine the canonical Bloch Rayleigh-Schrödinger (RS) perturbation theory (generation of effective Hamiltonians) with a Brillouin-Wigner (BW) correction step.~\cite{delafosse2024two,bindech2024combining} This so-called RSBW method enables the iterative construction of zeroth-order states and their energy corrections, offering a more flexible and scalable alternative to traditional multistate methods. In its previous formulation, the RSBW approach relied on a systematic RS optimization of all reference states, applying a multi-step treatment when needed to refine individual zeroth-order functions. This approach proved effective on model Hamiltonians, revealing that strongly correlated or near-degenerate states may require multiple iterations to achieve optimal decoupling.

However, only a limited number of transitions are of spectroscopic interest
 in many molecular applications.
In such cases, a full optimization of all zeroth-order states, as required in the original RSBW formulation, becomes unnecessarily costly. To address this issue, we propose here a $selective$ extension of the RSBW method, designed to target only a subset of the low-energy spectrum. By freezing previously optimized states and restricting RS treatments to newly selected ones, this approach significantly reduces the computational overhead while preserving the accuracy and rapid convergence of the BW correction. The state-specific RSBW (SS-RSBW) method provides a more practical route for realistic electronic structure calculations.
The remainder of the paper is organized as follows. Section 2 outlines the general procedure underlying the selective RSBW method proposed in this work. In the following sections 3-5, the method is applied to the LiH and H$_4$ molecules, focusing on the lowest singlet states. Those systems serve as a realistic test cases to assess the performance and scalability of the approach.

\section{State-specific RSBW  scheme}
This section details the key steps of the SS-RSBW procedure. The method builds on the foundations of
RS and BW
perturbation theories, combining iterative effective Hamiltonian constructions with energy corrections tailored to individual electronic states. 
A key aspect of the approach is the progressive decoupling of selected zeroth-order states from a strategically chosen subset of the remaining zeroth-order states, enabling state-specific and accurate energy predictions while maintaining compact zeroth-order representations.
The methodology is introduced through the sequential treatment of the ground and excited states, starting from a reference 
electronic configurations basis 
{which is gradually refined} via a sequence of effective model space optimizations and perturbative corrections.

\subsection{Zeroth-order Hamiltonian and initial partitioning}

Let $\hat{H}^{(0)}$ denote the Hamiltonian of an arbitrary electronic system, expressed in a finite 
electronic configurations
orthonormal basis $\mathcal{B}^{(0)} = \{\vert \psi_{k}^{(0)} \rangle\}_{0 \leq k \leq N-1}$.  In this basis, the Hamiltonian is decomposed as:
\begin{equation}
\begin{aligned}
    \hat{H}^{(0)} &= \sum_{kl} H_{kl}^{(0)} \vert \psi_{k}^{(0)}\rangle \langle \psi_{l}^{(0)}\vert \\
    &= \underbrace{\sum_{k} H_{kk}^{(0)} \vert \psi_{k}^{(0)}\rangle \langle \psi_{k}^{(0)}\vert}_{\hat{H}_0^{(0)}} 
    + \underbrace{ \sum_{k \neq l} H_{kl}^{(0)} \vert \psi_{k}^{(0)}\rangle \langle \psi_{l}^{(0)}\vert}_{\hat{W}^{(0)}} \\
    &= \sum_{k} E^{(0)}_{k} \vert \psi_{k}^{(0)}\rangle \langle \psi_{k}^{(0)}\vert + \sum_{k \neq l} W_{kl}^{(0)} \vert \psi_{k}^{(0)}\rangle \langle \psi_{l}^{(0)}\vert.
\end{aligned}
\end{equation}

In this decomposition, $\hat{H}_0^{(0)}$ defines the zeroth-order system, which serves as a reference Hamiltonian, while $\hat{W}^{(0)} = \hat{H}^{(0)} - \hat{H}_0^{(0)}$ represents the correlation operator, collecting all off-diagonal contributions. $E_k^{(0)}$ is the zeroth-order eigenenergy associated with the eigenstate $\vert \psi_k^{(0)} \rangle$ of $\hat{H}_0^{(0)}$. Throughout, we assume the states are energy-ordered, i.e., $E_k^{(0)} \leq E_{k+1}^{(0)}$.
Our goal is to estimate the energy of the ground state as well as those of the $M$ lowest excited states of the system. The proposed approach proceeds iteratively: we first compute the ground state energy, followed by the first excited state, and so on up to the $M$-th excited state.
First, the reference wavefunction is constructed
by diagonalizing successive effective Hamiltonians derived from
the RS perturbation theory, a step we refer to as the RS treatment.
In the second step, this reference 
state is energy-corrected at second order using a BW perturbative scheme, which captures the residual contributions that are not included in the zeroth-order description.
We now detail both steps, starting with the construction of the ground-state reference.

\subsection{Ground-state optimization via RS treatments}
\label{sec:RS treatment}

In our procedure, a zeroth-order eigenstate $\vert \psi_k^{(0)} \rangle$ of the reference Hamiltonian $\hat{H}_0^{(0)}$ 
is deemed an optimal reference state
for approximating an
eigenstate of the full Hamiltonian $\hat{H}$ if the following condition is fulfilled:
\begin{equation}
    \rho_{kl} = \left | \frac{\langle \psi_k^{(0)} | \hat{W}^{(0)} | \psi_l^{(0)} \rangle}{E_k^{(0)} - E_l^{(0)}} \right | \leq \rho_{\rm min}, \quad \forall l \neq k,
    \label{eq:trshold}
\end{equation}
where $\rho_{\rm min}$ is a predefined threshold.

To construct an optimal zeroth-order state for the ground state, we begin by selecting $\vert \psi_0^{(0)} \rangle$ as the initial reference ($k=0$). This state will be referred to as the Zeroth-Order Candidate State (ZO-CS). 
All $\vert \psi_{\alpha}^{(0)} \rangle$ states for which 
$\rho_{0\alpha}$ (see Eq.~\ref{eq:trshold}) exceeds the threshold 
$\rho_{\rm min}$ are progressively collected to build the model space $P_0$.~\cite{bindech2024combining} 
Particular care is required, however, when constructing the second-order 
effective Hamiltonian defined in Eq.~(\ref{eq:effective-hamiltonian}). 
At second order, non-hermitian contributions may arise; to address this, 
we employ a symmetrization procedure in which the Hamiltonian matrix is 
averaged with its transpose. Remarkably, the resulting operator can be 
directly related to the canonical van Vleck QDPT 
expansion.~\cite{lang2020combination,van1929sigma,shavitt1980quasidegenerate} 
In addition, some $Q$-space states $\vert \psi_{\beta}^{(0)} \rangle$ may 
interact strongly with selected $\vert \psi_{\alpha}^{(0)} \rangle$ states. 
To account for this, the model space is further enriched by including the 
most relevant of these configurations, selected according to a larger 
threshold $\rho'_{\rm min} > \rho_{\rm min}$.
Although these additional states are not directly coupled to the ZO-CS, including them in the model space improves the conditioning of the effective Hamiltonian and thus ensures numerical stability during RS optimization.

The diagonalization of the effective Hamiltonian yields a new set of orthonormal eigenstates $\{\vert \psi_{\alpha}^{\rm RS} \rangle\}$, 
and an updated basis:
\[
\mathcal{B}^{(1)} = \{ \vert \psi_{k}^{(1)} \rangle \} = \{\vert \psi_{\alpha}^{\rm RS} \rangle\} \cup \{\vert \psi_{\beta}^{(0)} \rangle\},
\]
where the superscript indicates that a single RS treatment has been performed.
The objective of this step is to concentrate part of the electronic effects
and to improve the description of the ZO-CS. 

The zeroth-order Hamiltonian is then updated as:
\begin{equation}
\begin{aligned}
\hat{H}_0^{(1)} &= \sum_{k} E_{k}^{(1)} \vert \psi_{k}^{(1)} \rangle \langle \psi_{k}^{(1)} \vert \\
&= \underbrace{\sum_{\alpha} E_{\alpha}^{\rm RS} \vert \psi_{\alpha}^{\rm RS} \rangle \langle \psi_{\alpha}^{\rm RS} \vert}_{\text{updated}} + \underbrace{\sum_{\beta} E_{\beta}^{(0)} \vert \psi_{\beta}^{(0)} \rangle \langle \psi_{\beta}^{(0)} \vert}_{\text{unchanged}}.
\end{aligned}
\end{equation}
The new coupling operator is given by $\hat{W}^{(1)} = \hat{H}^{(1)} - \hat{H}_0^{(1)}$, where $\hat{H}^{(1)}$ stands for the representation of the Hamiltonian in the $\mathcal{B}^{(1)}$ basis.
Note that the matrix structure of the Hamiltonian remains unchanged within the orthogonal space, whereas  the matrix elements involving the model space are updated.
Maintaining the energy ordering 
convention,
the new ZO-CS is $\vert \psi_0^{(1)} \rangle$, and the
{decoupling condition~\eqref{eq:trshold}
is re-examined to generate
an updated
model space $P_0$.}
The procedure is repeated until 
an optimized ZO-CS $\vert \psi_0^{(n_0^{\rm RS})} \rangle$ is obtained, where $n_0^{\rm RS}$ is the number of
required RS steps.
This state serves as the zeroth-order reference for the ground-state energy.

\subsection{Brillouin–Wigner energy correction}

After $n_0^{\rm RS}$ RS steps, the zeroth-order ground-state energy
is then refined using a second-order Brillouin–Wigner (BW) perturbation correction to reach SS-RSBW energy values. The reference wavefunction is $\vert \psi_0^{(n_0^{\rm RS})} \rangle$ 
and the zeroth-order energy given by
$E_0^{(n_0^{\rm RS})} = \langle \psi_{0}^{(n_0^{\rm RS})} | \hat{H}_0^{(n_0^{\rm RS})} | \psi_{0}^{(n_0^{\rm RS})} \rangle$.
Assuming that the exact BW energy expansion
can be approximated by its second-order value
$E_0^{\mbox{\tiny SS-RSBW}}$,
the BW energy correction reads\cite{lindgren2012atomic}:
\begin{align}
E_0^{\mbox{\tiny SS-RSBW}} &= E_0^{(n_0^{\rm RS})} 
+ \langle \psi_{0}^{(n_0^{\rm RS})} | \hat{W}^{(n_0^{\rm RS})} | \psi_{0}^{(n_0^{\rm RS})} \rangle \nonumber \\
&\quad + \sum_{j \neq 0} \frac{ \left| \langle \psi_{0}^{(n_0^{\rm RS})} | \hat{W}^{(n_0^{\rm RS})} | \psi_j^{(n_0^{\rm RS})} \rangle \right|^2 }{E_0^{\mbox{\tiny SS-RSBW}} - E_j^{(n_0^{\rm RS})}}.
\label{eq:BW_general}
\end{align}

This equation is either solved  iteratively, 
or by setting $E_0^{\mbox{\tiny SS-RSBW}} \approx E_0^{(n_0^{\rm RS})}$ in the energy denominators.

\subsection{Iterative extension to excited states}

The same two-step strategy can be sequentially applied to describe excited states. 
To compute the energy of the first excited state, we start from the optimized zeroth-order basis obtained in the previous step, denoted $\mathcal{B}^{(n_0^{\mbox{\tiny RS}})}$. In this basis, the zeroth-order Hamiltonian is rewritten as follows:
\begin{align}
    \hat{H}^{(n_0^{\mbox{\tiny RS}})}_0 &= \overbrace{E^{(n_0^{\mbox{\tiny RS}})}_{0} \, \vert \Psi^{(n_0^{\mbox{\tiny RS}})}_{0} \rangle \langle \Psi^{(n_0^{\mbox{\tiny RS}})}_{0} \vert}^{\text{Optimized}} \nonumber \\
    &+ \underset{k \geq 1}{\sum} E^{(n_0^{\mbox{\tiny RS}})}_{k} \, \vert \Psi^{(n_0^{\mbox{\tiny RS}})}_{k} \rangle \langle \Psi^{(n_0^{\mbox{\tiny RS}})}_{k} \vert
\label{eq:First-part}
\end{align}

This partitioning isolates the already optimized part of the spectrum from the remaining states. The former is referred to as the \textit{Optimized Zeroth-Order space} (OZO-space). At this stage, the OZO-space is restricted to the optimized zeroth-order ground state function $\vert \psi_{0}^{(n_0^{\rm RS})} \rangle$.
Let us emphasize that
the states within the OZO-space are treated as perturbers in the subsequent RS treatments. That is, they are no longer modified during the optimization of the zeroth-order excited states. 

The ZO-CS
for targeting the first excited state is the lowest-lying state outside the OZO-space, namely $\vert \psi_{1}^{(n_0^{\rm RS})} \rangle$. 
Following the previous strategy with identical 
$\rho_{\rm min}$ and $\rho'_{\rm min}$ thresholds, the model space $P_1$ is constructed from the 
$\vert \psi_l^{(n_0^{\rm RS})} \rangle$ 
states with $l \geq 2$.
After each RS step, the energies and wavefunctions of the optimized states in the OZO-space are kept frozen, while the remaining ones are energy reordered as needed.
The RS optimization is repeated until a new lowest-energy state, orthogonal to the OZO-space, is identified. The resulting optimized state is denoted $\vert \psi_{1}^{(m_1^{\rm RS})} \rangle$, where $m_1^{\rm RS} = n_0^{\rm RS} + n_1^{\rm RS}$ corresponds to the cumulative number of RS steps. This new state is transfered to the OZO-space, and its energy $E^{\mbox{\tiny RSBW}}_{1}$ is computed using the second-order BW correction as defined in Eq.~\eqref{eq:BW_general}.
This iterative procedure is applied to construct successively the $M$ lowest
eigenstates of the full Hamiltonian. The newly $i$-th zeroth-order state computed after $n_i^{\rm RS}$ RS steps is added to the OZO space.
This sequential decoupling approach ensures that each targeted state is treated independently, while maintaining the structure of OZO-space states. 
However, it should be noted that the algorithm does not guarantee that the order in which states are optimized may differ from the exact one. For instance, the second excited state may occasionally be obtained before the first one. 
The final set of SS-RSBW energies is systematically reordered in ascending energy at the end of the calculation. A compact summary of the full sequence of operations defining the SS-RSBW scheme used in this work is presented in Algorithm~\ref{code:RSBW_pseudoalgo}.

\begin{algorithm}[h!] 
\caption{SS-RSBW procedure for computing the energies of  $M$ targeted states.
In practice, the procedure allows to generate the lowest states starting from the Hartree-Fock $\hat{H}^{(0)}$ Hamiltonian.}

\begin{algorithmic}[1]

\State \textbf{Step 1: Initialization}
\State Set thresholds: $\rho_{\rm min} \gets \text{e.g., } 0.1$, \quad $\rho'_{\rm min} \gets \text{e.g., } 0.5$
\State Define $\hat{H}_0^{(0)}$ such that $(H_0^{(0)})_{ij} = \delta_{ij} H_{ij}^{(0)}$
\State Set $\hat{W}^{(0)} \gets \hat{H}^{(0)} - \hat{H}_0^{(0)}$
\State Assign  $\mathcal{B}^{(0)}= \{ \ket{i} \}$ to the zeroth-order eigenvectors of  $\hat{H}_0^{(0)}$ with corresponding eigenvalues $E_i^{(0)} \gets \bra{i} \hat{H}_0^{(0)} \ket{i}$
\vspace{0.2em}
\State Set $i \gets 0$ \Comment{Index of the targeted state}
\State Set $m_i^{\rm RS} \gets 0 $ \Comment{Cumulative number of RS iterations for state $i$}

\vspace{0.5em}
\Loop{ over $i = 0, \dots, M-1$} 
  \State \textbf{Step 2: Iterative RS treatment}
  \State Build model space $P_i = \left\{|\psi_\alpha^{(m_i^{\rm RS})}\rangle\right\}$ 

  \While{dim$(P_i) > 1$}
    \State Build orthogonal $Q_i$-space
    \State Build and symmetrize $\hat{H}_{\rm eff}^{(2)}$
    \State $\left\{E_\alpha\right\}\gets$ Diagonalize $\hat{H}_{\rm eff}^{(2)}$
    \State $m_i^{\rm RS} \gets m_i^{\rm RS} + 1$
    \State Update $\mathcal{B}^{( m_i^{\rm RS})}$, $\hat{H}^{(m_i^{\rm RS})}$

     \State Update  $\langle \psi_\alpha^{(m_i^{\rm RS})}| \hat{H}_0^{(m_i^{\rm RS })} |\psi_\alpha^{(m_i^{\rm RS})}\rangle \gets  \left\{ E_\alpha\right \}$
     \State Set $\hat{W}^{(m_i^{\rm RS})} \gets \hat{H}^{(m_i^{\rm RS})} - \hat{H}_0^{(m_i^{\rm RS})}$
    \State Sort basis  states in ascending energy
    \State Build $P_i$ in the new basis $\mathcal{B}^{( m_i^{\rm RS})}$
  \EndWhile
  \vspace{0.5em}
  \State \textbf{Step 3: BW correction}
  \State Compute $E^{\mbox{\tiny RSBW}}_i$ using Eq.~(\ref{eq:BW_general2})
  \vspace{0.7em}
  \State Add $|\psi_i^{(m_i^{\mbox{\tiny RS}})}\rangle$ to the OZO-space 
  \State Set $i \gets i+1$,  $m_i^{\rm RS} \gets  m_{i-1}^{\rm RS}$ 
\EndLoop
\State Sort SS-RSBW energies in ascending energy order

\end{algorithmic}
\label{code:RSBW_pseudoalgo}
\end{algorithm}

\section{Numerical tools}

We numerically tested our approach on two prototypical molecular systems, LiH and H$_4$ ring, using STO-3G atomic orbitals (AOs): specifically, the $1s$ orbital for hydrogen and the $1s$, $2s$, and $2p$ orbitals for lithium.
The latter are used to generate canonical molecular orbitals (MOs) via Restricted Hartree-Fock (RHF) calculations 
implemented in the \texttt{Psi4} package.
~\cite{smith2020psi4}
Then, the \texttt{QuantNBody} package~\cite{yalouz2022quantnbody,codeQuantNBody} allows to transform the one- and two-electron integrals from the AOs to the MOs basis, generate the Slater determinants basis set, and construct the electronic Hamiltonian matrix.
To isolate the singlet subspace, a configuration state functions (CSFs) basis is constructed from spin-adapted linear combinations of Slater determinants. This is done by building and diagonalizing the \(\hat{S}^2\) operator over the determinants space within \texttt{QuantNBody}, yielding an orthonormal spin-adapted basis. The Hamiltonian is then transformed into this basis, becoming block-diagonal,
and the singlet CSFs define the 
zeroth-order basis \( \mathcal{B}^{(0)} \) used throughout the procedure.

\section{Application to lithium hydride}

The proposed SS-RSBW method was first applied 
to compute the low-lying singlet electronic states of the lithium hydride (LiH) molecule. Specifically, we focused on the ground and the first two excited singlet states, 
as a function of the internuclear distance \( R \).
The FCI space is spanned by
the Slater determinants arising from distributing four electrons in twelve spin-orbitals, leading to $N = \binom{12}{4} = 495$ 
determinants.
The FCI ground-state energy profile exhibits a minimum at \( R_{\rm eq} = 1.55~\AA \), showing a small deviation from the reported equilibrium bond distance of \( 1.60~\AA \). Despite this discrepancy, the STO-3G basis set was retained to generate the reference FCI calculations.
To investigate the evolution of the electronic structure across both compressed and stretched bonding regimes, the internuclear distance \( R \) is varied from \( 0.75 \) to \( 2.20~\AA \). 
The zeroth-order basis \( \mathcal{B}^{(0)} \)
is defined by the \textbf{$N=105$}
singlet CSFs.

\subsection{ZO-CSs optimization: RS treatments}

The optimization of the ZO-CSs was carried out using the RS 
scheme introduced in Section~\ref{sec:RS treatment}. The selection of states included in the model space is governed by the two thresholds \( \rho_{\rm min} = 0.1 \) and \( \rho'_{\rm min} = 0.5 \), used across all targeted states.

\begin{figure}[H]
\centering
\includegraphics[width=8cm]{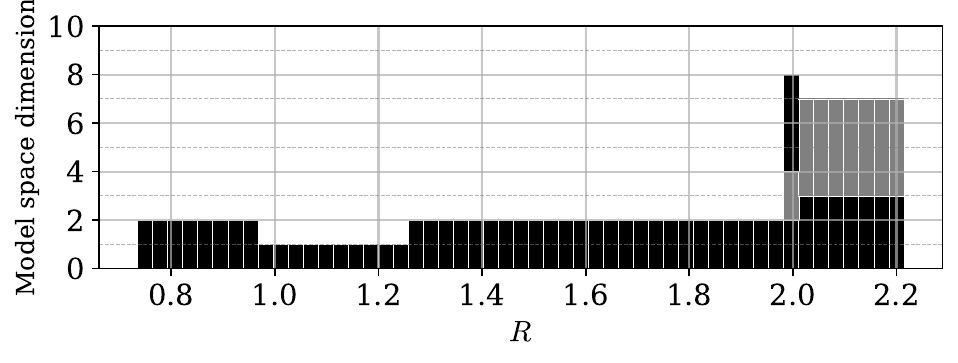}
\caption{\footnotesize  $P_0$ model space dimensions used for the optimization of the ZO-CS for the ground state as a function of the internuclear distance  $R$ ($\AA$). Bar stackings indicate the dimensions of the successive model spaces $P_0$ at each RS step. 
}
\label{fig:model-space-0}
\end{figure}

\uline{Ground state}: The initial ZO-CS for the ground state was selected as the lowest-lying CSF. 
The dimensions of the model spaces during the successive RS treatments used to optimize this reference state are shown in Figure~\ref{fig:model-space-0}.

\begin{figure}[h]
\centering
\includegraphics[width=8.2cm]{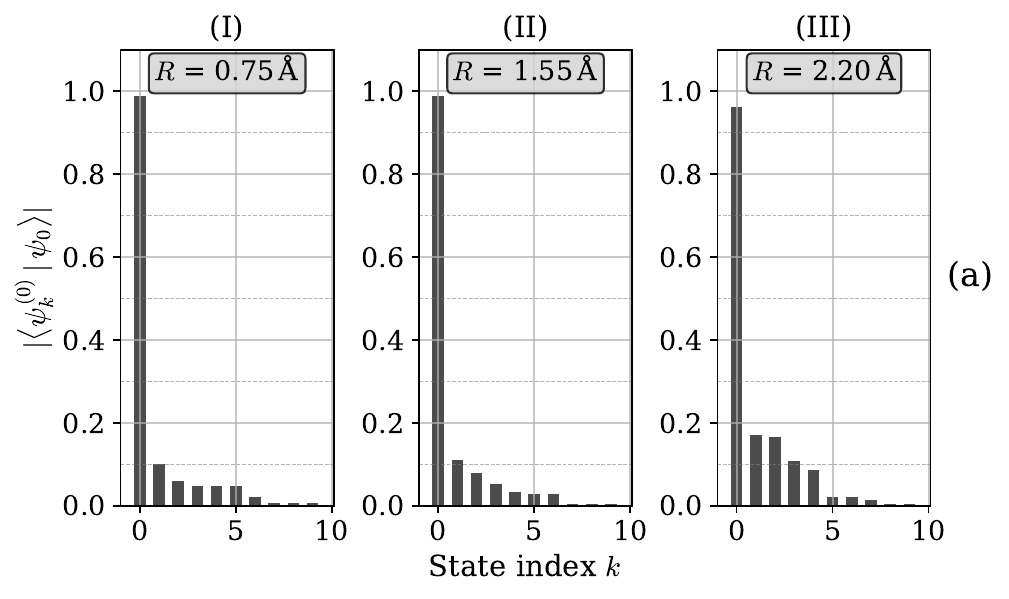}
\includegraphics[width=8.2cm]{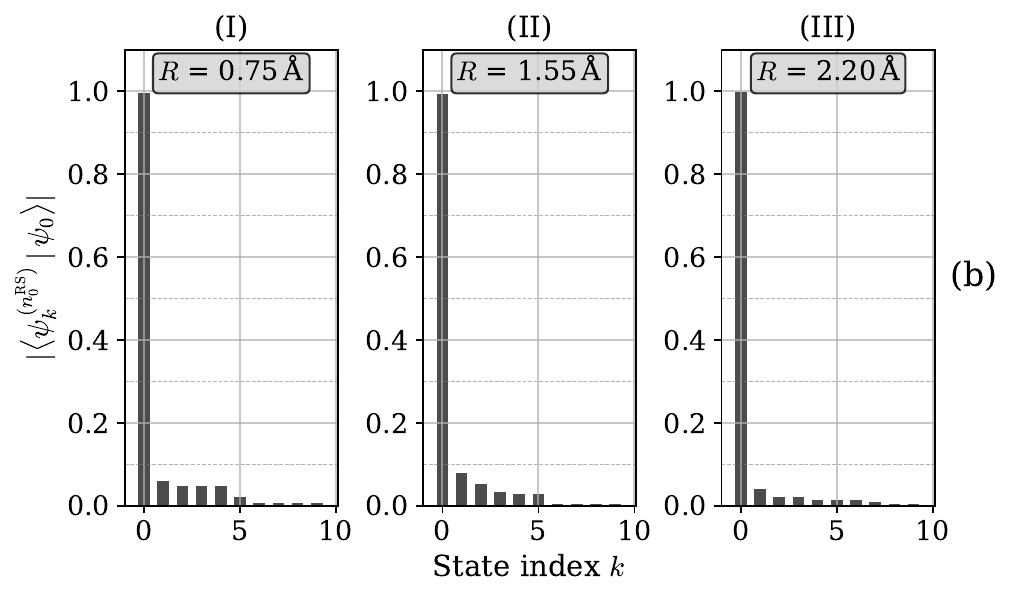}
\caption{\footnotesize Decomposition of the exact ground state on (a) the initial CSFs basis \( \mathcal{B}^{(0)} \), and (b) the optimized basis \( \mathcal{B}^{(n_0^{\rm RS})} \) for three internuclear
distances $R$ ($\AA$). The projection onto the ZO-CS is indexed by \( k=0 \), and others are ordered by descending absolute value. Three regimes are shown: (I) compressed, (II) equilibrium, and (III) stretched geometries.}
\label{fig:decomp-GS}
\end{figure}

For internuclear distances \( R < 2.0~\AA \), the model
space dimension is less than two, and the number of RS steps is
$m_0^{\rm RS} = 1$. As expected, the dimension of the model space as
well as $m_0^{\rm RS}$ increase in the bond-stretched regime \( R > 2.0~\AA \) where static correlation effects must be included.
To assess the impact of RS optimization, the decompositions of the exact ground-state FCI wavefunction
are compared for the initial and optimized zeroth-order bases in  Figure~\ref{fig:decomp-GS}.

The decomposition of the ground-state wavefunction in the initial CSF basis (Figure~\ref{fig:decomp-GS}-a) exhibits a strong dominance of the 
{ZO-CS}
which is primarily composed of the RHF determinant. This observation is consistent with the well-established accuracy of the RHF description for ground-state properties in weakly correlated regimes~\cite{helgaker2013molecular}. However, as the LiH bond length 
$R$ increases, the wavefunction acquires a pronounced multiconfigurational character (i.e., static correlation), as evidenced by the growing number of significant contributions from other CSFs in
the $\mathcal{B}^{(0)}$ basis (Figure~\ref{fig:decomp-GS}-a(III)).
The picture is significantly modified in the 
$\mathcal{B}^{(n_0^{\mbox{\tiny RS}})}$ basis.
As seen in Figure \ref{fig:decomp-GS}-b, the projection onto the optimized ZO-CS dominates
as soon as  the  $\mathcal{B}^{(n_0^{\mbox{\tiny RS}})}$ basis is constructed. The sequence of RS treatments in this region systematically incorporates the most relevant configurational components needed for an accurate zeroth-order state that offers a significantly improved description of the ground state compared to the initial CSF-based ZO-CS.

\begin{figure}[H]
\centering
\includegraphics[width=8cm]{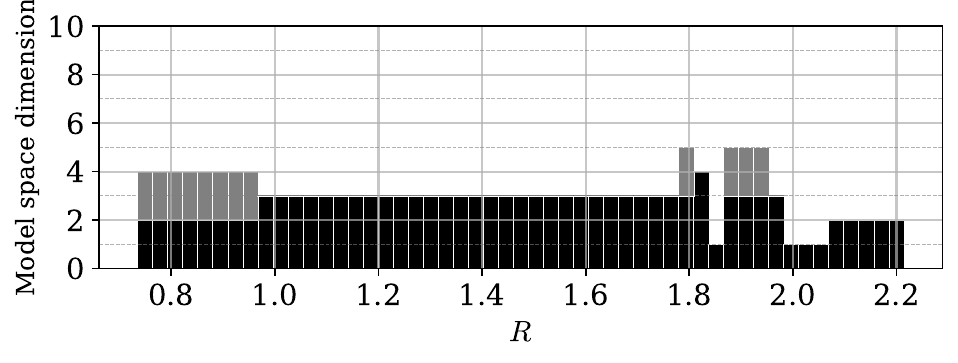}
\caption{\footnotesize $P_1$ model spaces dimensions  used throughout the optimization of the ZO-CS for the first excited singlet state
as a function of $R$ ($\AA$). For further details, refer to the caption of Figure \ref{fig:model-space-0}.}
\label{fig:model-space-1}
\end{figure}

\begin{figure}[H]
\centering
\includegraphics[width=8.2cm]{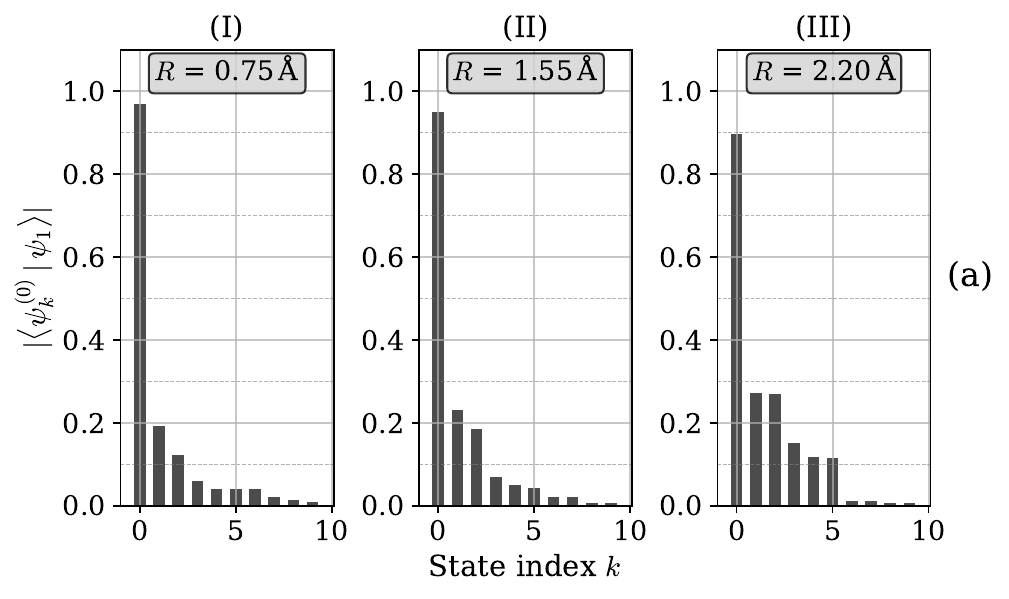}
\includegraphics[width=8.2cm]{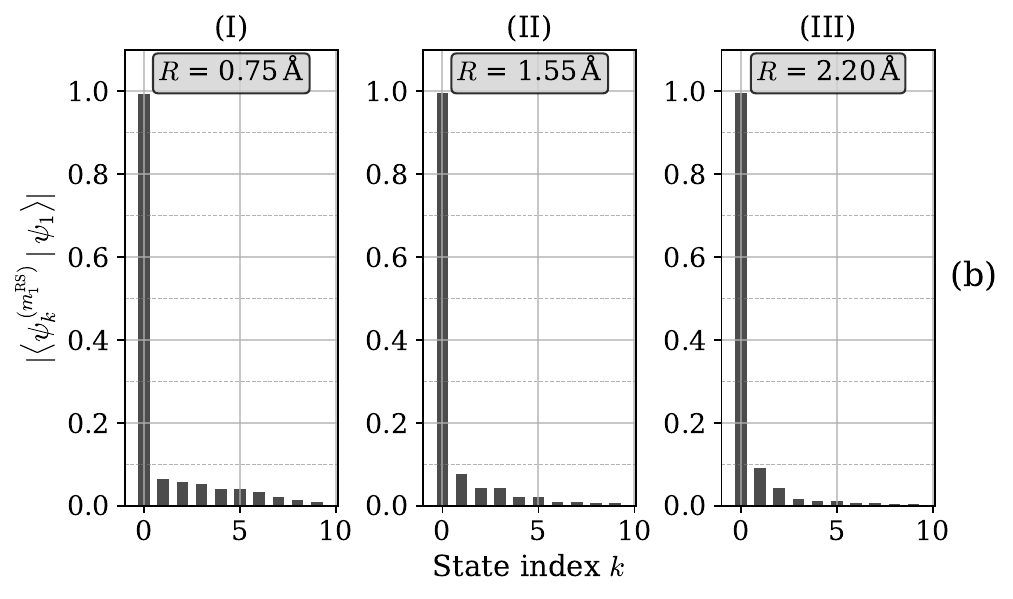}
\caption{\footnotesize Decomposition of the exact first excited singlet state on (a) the initial CSF basis \( \mathcal{B}^{(0)} \), and (b) the optimized basis \( \mathcal{B}^{(m_1^{\rm RS})} \). In the optimized basis, the projection onto the candidate state is indexed by $k=0$, while the projections onto the other states are ordered in decreasing absolute value. In the CSFs basis, projections are ordered by decreasing absolute value. For further details, refer to the caption of Fig.~\ref{fig:decomp-GS}. }
\label{fig:decomp-1st}
\end{figure}

\uline{First excited singlet state}: A similar analysis was carried out for the first excited singlet state. The initial ZO-CS is selected from the updated basis \( \mathcal{B}^{(n_0^{\rm RS})} \) as the lowest-energy state orthogonal to the OZO-space, namely \( \vert \psi_{1}^{(n_0^{\rm RS})} \rangle \). The associated model space dimensions $P_1$ and wavefunction decompositions are presented in Figures~\ref{fig:model-space-1} and~\ref{fig:decomp-1st}, respectively.

First, the model space dimensions $P_1$ involved in the successive RS treatments 
remain rather small 
whatever the internuclear distance (see Figure \ref{fig:model-space-1}). 
Interestingly, the starting ZO-CS $\vert \psi_{1}^{(n_0^{\rm RS})} \rangle$, benefits from the refinements performed through the optimization of the ground state. 
Despite its higher multiconfigurational character 
in the \(  \mathcal{B}^{(0)} \) basis set 
(see Figure~\ref{fig:decomp-1st}, panel (a)), the computational effort required for its optimization is significantly reduced when 
the 
\textbf{\( \mathcal{B}^{(n_0^{\rm RS})} \)} basis set is used. 
Finally, Figure \ref{fig:decomp-1st}-b shows that the first excited singlet state exhibits a clear dominant projection on the ZO-CS, illustrating the ability of the RS treatment to effectively concentrate the essential electronic information into the selected reference state.

\begin{figure}[H]
\centering
\includegraphics[width=8cm]{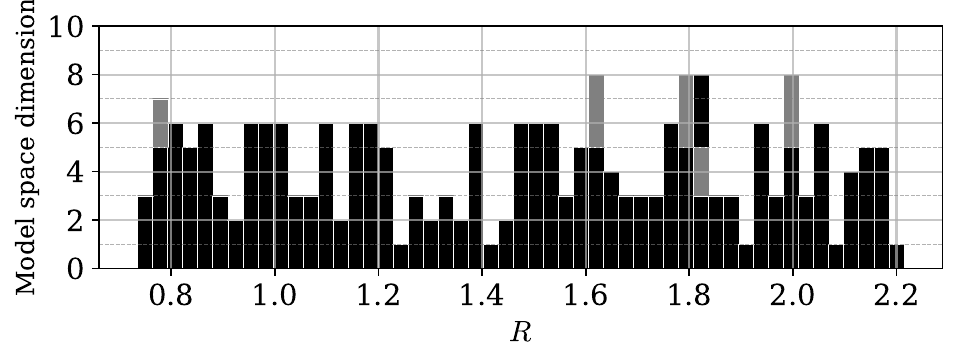}
\caption{\footnotesize Model spaces dimensions $P_2$ used throughout the optimization of the ZO-CS for the second excited singlet state. For further details, refer to the caption of Figure \ref{fig:model-space-0}.}
\label{fig:model-space-2}
\end{figure}

\uline{Second excited singlet state}: The ZO-CS for the second excited singlet state was optimized using the same sequential RS strategy, with \( \vert \psi_{2}^{(m_1^{\rm RS})} \rangle \) (\( m_1^{\rm RS} = n_0^{\rm RS} + n_1^{\rm RS} \)) selected as the initial reference state. {As highlighted for the first excited state, the preceding RS treatments may contribute to a pre-conditioning of the state \( \vert \psi_{2}^{(m_1^{\rm RS})} \rangle \), thereby reducing the computational effort required for its optimization. This is reflected in the dimensions of the model spaces $P_2$ which remain relatively small,
as illustrated in Figure \ref{fig:model-space-2}}. The  observed behavior {for the optimization of the ZO-CS for the second excited singlet state} closely parallels that of the first excited state: the RS treatments reliably yield a dominant component in the optimized state, even in strong correlation regimes. 

These results underscore the efficiency of the RS procedure in capturing multiconfigurational effects within a single optimized vector, thereby enabling a compact yet accurate representation of both ground and excited electronic states across a broad range of correlation regimes.

\begin{figure}[H]
\centering
\includegraphics[width=8.2cm]{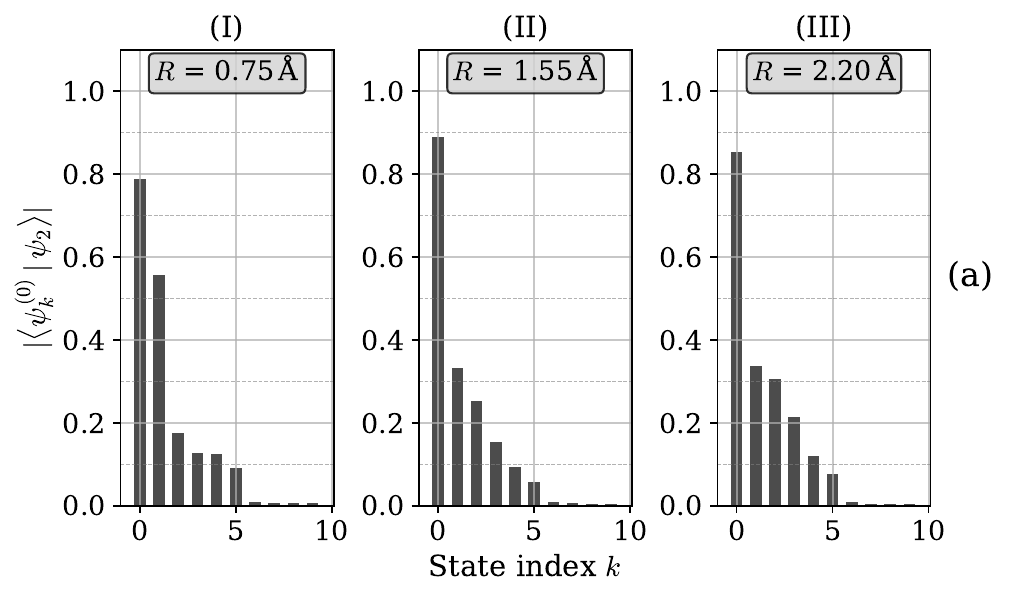}
\includegraphics[width=8.2cm]{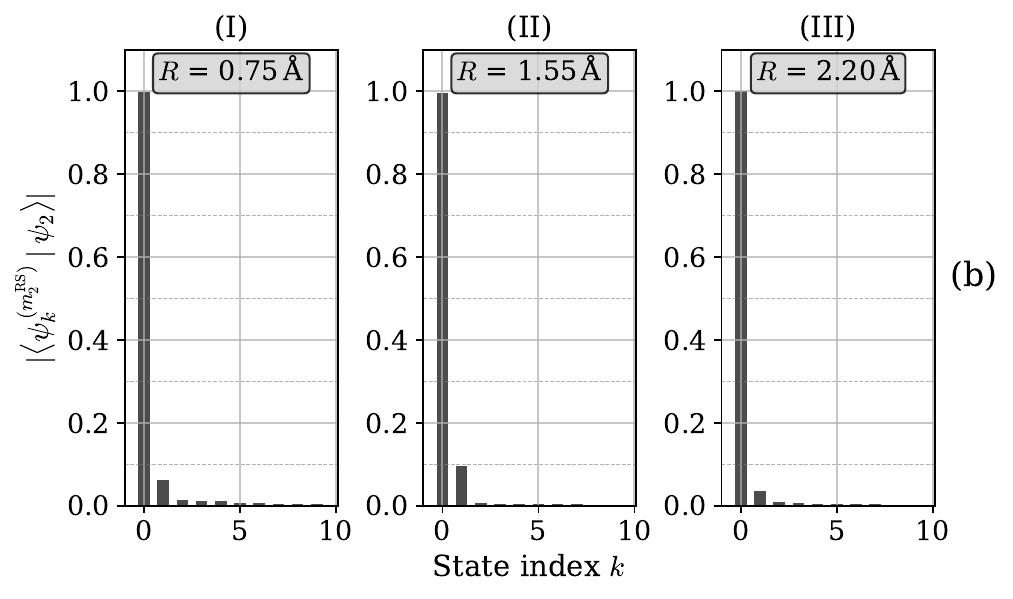}
\caption{\footnotesize Decomposition of the exact first excited singlet state on (a) the initial CSFs basis \( \mathcal{B}^{(0)} \), and (b) the optimized basis \( \mathcal{B}^{(m_2^{\rm RS})} \), where \( m_2^{\rm RS} \) is the total number of RS steps. In the optimized basis, the projection onto the candidate state is indexed by $k=0$, while the projections onto the other states are ordered in decreasing absolute value. In the CSF basis, projections are ordered by decreasing absolute value. For further details, refer to the caption of Figure \ref{fig:decomp-GS}.}
\label{fig:decomp-2nd}
\end{figure}

\subsection{Energy evaluation via BW correction}

The  SS-RSBW energies of the singlet states were computed by applying second-order BW corrections to the zeroth-order energies of the optimized states \( \{ \vert \psi_k^{(m_k^{\rm RS})} \rangle \} \). The corrected energy \( E_k^{\mbox{\tiny SS-RSBW}} \) is defined by:
\begin{align}
E_k^{\mbox{\tiny SS-RSBW}} &= E_k^{(m_k^{\rm RS})} 
+ \langle \psi_{k}^{(m_k^{\rm RS})} | \hat{W}^{(m_k^{\rm RS})} | \psi_{k}^{(m_k^{\rm RS})} \rangle \nonumber \\
&\quad + \sum_{j \neq k} \frac{ \left| \langle \psi_{k}^{(m_k^{\rm RS})} | \hat{W}^{(m_k^{\rm RS})} | \psi_j^{(m_k^{\rm RS})} \rangle \right|^2 }{E_k^{\mbox{\tiny SS-RSBW}} - E_j^{(m_k^{\rm RS})}}.
\label{eq:BW_general2}
\end{align}

where \( E_k^{(m_k^{\rm RS})} = \langle \psi_{k}^{(m_k^{\rm RS})} | \hat{H}_0^{(m_k^{\rm RS})} | \psi_{k}^{(m_k^{\rm RS})} \rangle \).

\begin{figure}[H]
\centering
\includegraphics[width=8cm]{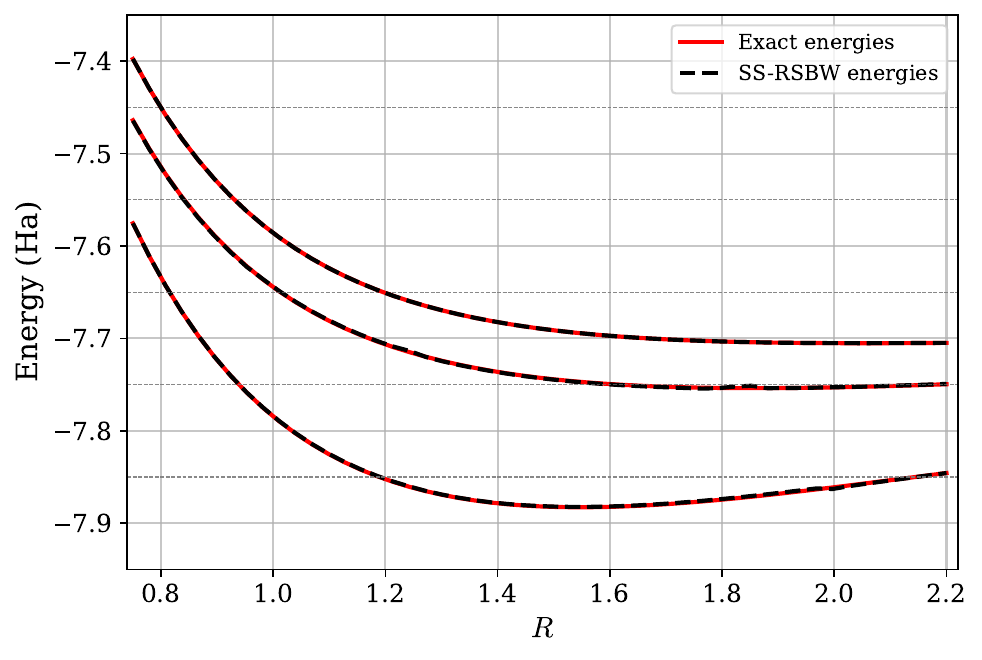}
\caption{\footnotesize SS-RSBW energies \( \{E^\text{SS-RSBW}_{0\text{–}2}\} \) of the three lowest singlet states of LiH as a function of the internuclear distance \( R \). Exact FCI results \( \{E^{\mathrm{exact}}_{0\text{–}2}\} \) are shown for comparison. 
}
\label{fig:E_RSBW LiH}
\end{figure}

The quality of the computed energy levels obtained from both RS and SS-RSBW methods, was assessed using the root-mean-square (RMS) error, defined as

\[
\Delta^\text{method}
= \sqrt{\frac{1}{N_R} \sum_{i_R=1}^{N_R} \left( E^\text{method}_{i_R} - E^{\rm FCI}_{i_R} \right)^2},
\]
where the upperscript "method" is either RS or SS-RSBW, and  \( N_R = 51\) denotes the number of grid points used to sample the internuclear distance \( R \). 
The RMS errors $\Delta^{\rm RS}$ and  
$\Delta^\text{SS-RSBW}$
for the three computed singlet energy levels are reported in Table~\ref{tab:errors}.
\renewcommand{\arraystretch}{1.2} 
\begin{table}[H]  
 \begin{center}
  \begin{tabular}{c|c|c|c}
    \hline 
    Energy Level & \( E_0 \) & \( E_1 \) & \( E_2 \) \\
    \hline \hline
    \( \Delta^{\rm RS} \) (mHa) & \( 1.0 \) & \( 2.3  \) & \( 0.1  \) \\
    \hline 
    \( \Delta^\text{SS-RSBW} \) (mHa) & \( 0.07 \) & \( 0.08 \) & \( 0.003 \) \\
    \hline 
    
  \end{tabular}
  \caption{Root-mean-square  errors (mHa) for the three lowest singlet energy levels of LiH. 
The first row reports errors from RS treatments, while the second row corresponds to the full SS-RSBW results. 
Reference values are obtained from FCI calculations. All computations are performed using the STO-3G AO basis set.
  }
  \label{tab:errors}
\end{center}
\end{table}

Table~\ref{tab:errors} suggests that the RS treatments already provide remarkably accurate zeroth-order energies over the bond-length range, reaching spectroscopic accuracy (RMS 
$\sim$
1 mHa).
This result underscores the robustness of the underlying partitioning and confirms that the reference states are both accurate and systematically improvable. Building on this, the full SS-RSBW scheme further reduces the errors: for all three singlet states, the RMS deviations from the exact FCI energies remain below 0.1 mHa.
The algorithm converges rapidly, requiring less than five iterations to solve Eq.~\eqref{eq:BW_general2} for all geometries. This highlights both the quality of the RS-optimized zeroth-order states and the overall efficiency of the scheme for describing low-lying singlet states in correlated electronic systems. Besides, the enhanced convergence suggests that higher-order BW energy expansion might be included, shifting away the size-consistency drawback of the BW theory.


\section{Application to the H$_4$ ring}

\begin{figure}[H]
\centering
\includegraphics[width=5cm]{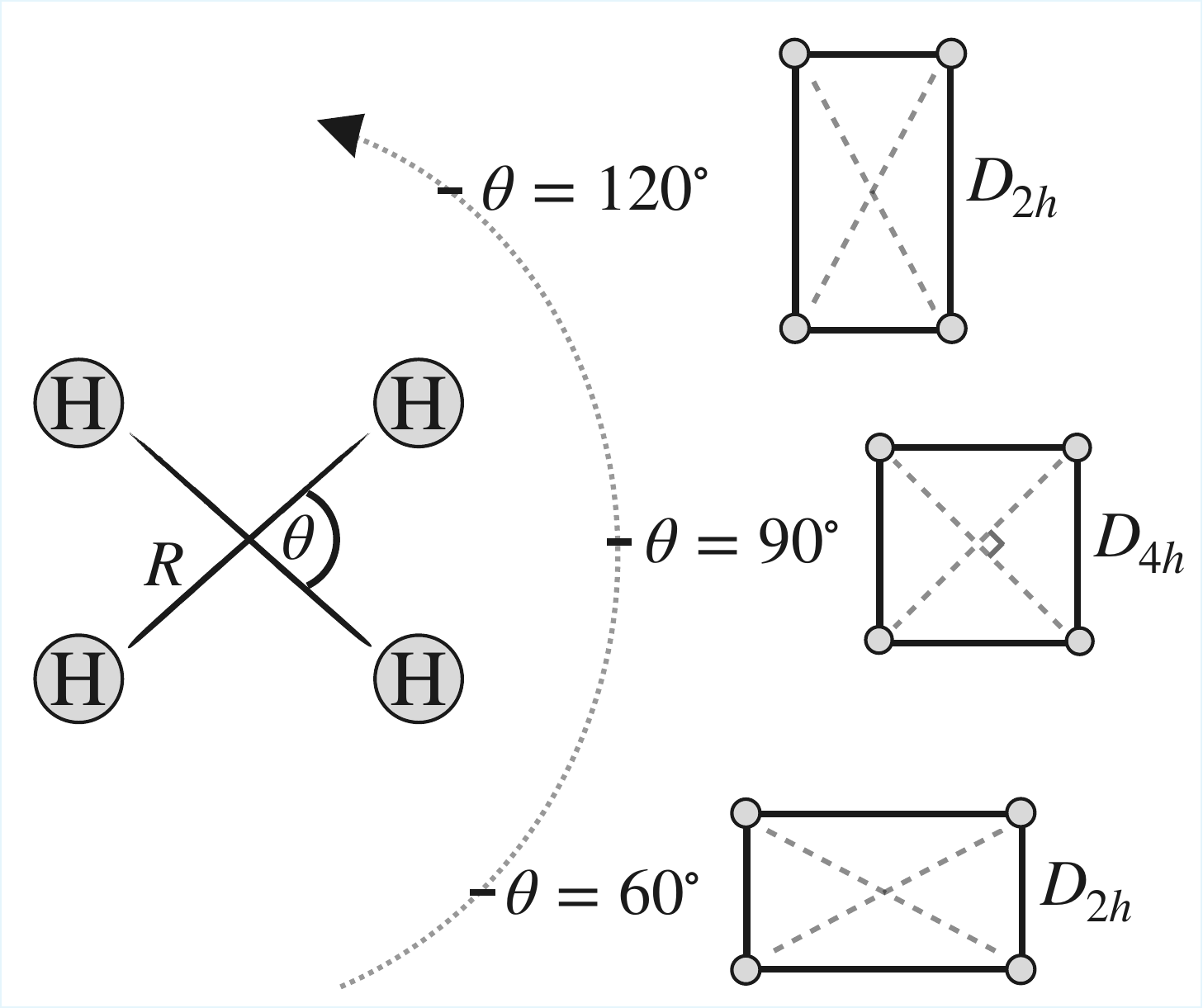}
\caption{\footnotesize Geometry of the ring H$_4$ model with angular parametrization. The radius $R$ is fixed at 1~\AA, and the angle $\theta$ varies from $60^\circ$ to $120^\circ$.}
\label{fig:H4-scheme}
\end{figure}
This section addresses the application of the SS-RSBW method to the H$_4$ ring model.
The system consists of four hydrogen atoms arranged on a circle of fixed radius $R=1.0\ \text{\AA}$, 
while keeping a C$_2$ axis
with a characteristic  $\theta\in[60^\circ,120^\circ]$ angle
(see Figure~\ref{fig:H4-scheme}).
This controlled deformation of the system provides a simple yet stringent testbed to probe the emergence and removal of orbital degeneracies associated with the transition between the $D_{2h}$ and $D_{4h}$ point groups. Such geometries induce strong static correlation effects that are poorly described by single-reference methods, making H$_4$ a classical benchmark for assessing the robustness of multireference approaches \cite{boguslawski2016analysis,marie2021variational,van2000benchmark,limacher2014influence}.
The FCI space built on four electrons in eight spin-orbitals comprises $\binom{8}{4}=70$ Slater determinants combined into 20 spin-adapted singlet CSFs.

\subsection{ZO-CS optimization: RS treatments}
   
The optimization of ZO-CSs for ground and
excited states through RS treatments follows the procedure presented earlier, using the thresholds values $\rho_{\min}=0.05$ and $\rho'_{\min}=0.2$.

\underline{Ground state:} 
\begin{figure}[h]
\centering
\includegraphics[width=8cm]{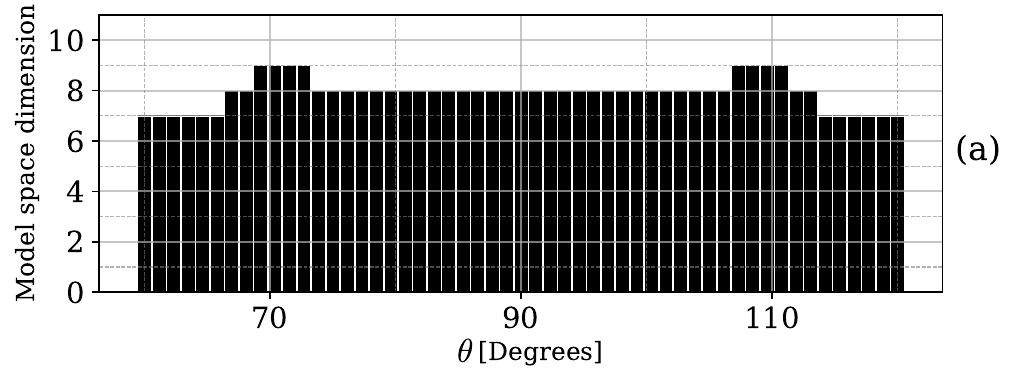}
\includegraphics[width=6.3cm]{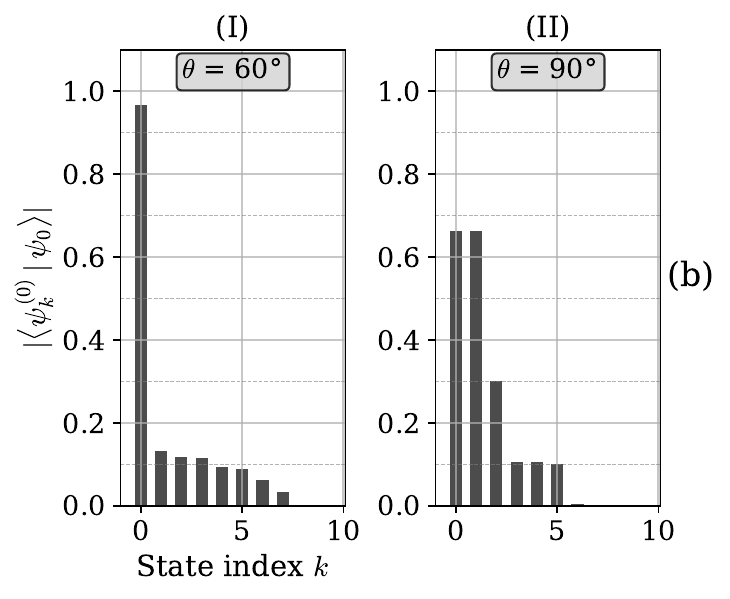}
\includegraphics[width=6.3cm]{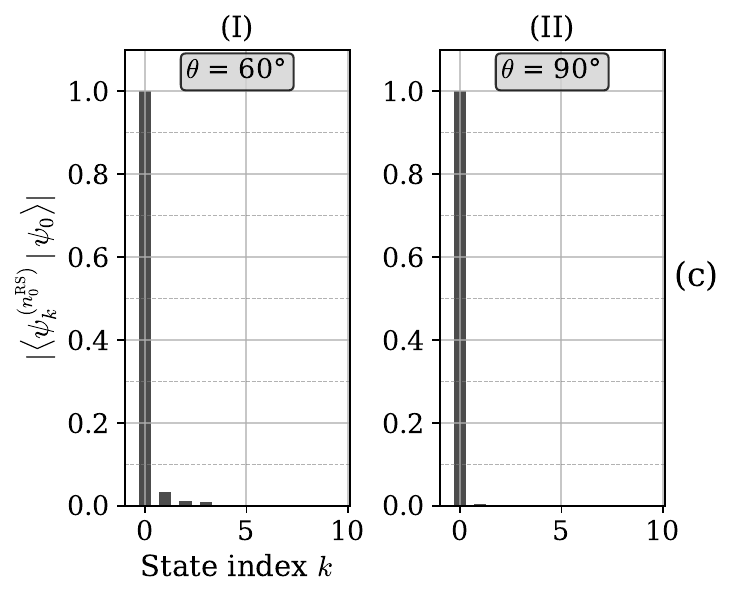}
\caption{\footnotesize Optimization of the ZO-CS for the ground singlet state: (a) dimensions of the model spaces $P_0$ as a function of the angle $\theta$; (b) and (c) decompositions of the FCI ground state over the initial CSF basis \( \mathcal{B}^{(0)} \),  and the optimized zeroth-order basis \( \mathcal{B}^{(n_0^{\rm RS})} \), respectively, shown for (I) rectangular and (II) square geometries. The projection onto the ZO-CS is indexed by \( k=0 \), and others are ordered by descending absolute values.}

\label{fig:H4-GS}
\end{figure}
As seen in Figure~\ref{fig:H4-GS}-a, a single iteration
($n_0^{\rm RS} = 1$)
is necessary to  optimize the ZO-CS  whatever the 
$\theta$ value. The strong multi-configurational character is reflected by the dimension of the model space that remains relatively large as compared to the 
dimension of the full CSFs basis set. 
Then, special attention was dedicated to the 
 rectangular $\theta = 60^\circ$ (I) and 
square $\theta = 90^\circ$ (II) geometries.
As seen in Figure~\ref{fig:H4-GS}-b, the decompositions of the ground state 
over \( \mathcal{B}^{(0)} \) reveal significant contributions arising from several CSFs,
with a more pronounced static correlation manifestation for
the square geometry (see Figure~\ref{fig:H4-GS}-b(II)).
The $D_{4h}$ symmetry induces orbital degeneracies which, in the singlet CSF basis, manifest as a broader expansion of the wavefunction.
Finally, Figure~\ref{fig:H4-GS}-c highlights a dominant projection of the FCI ground state onto the optimized ZO-CS for both geometries, thus confirming the reliability of the constructed reference state with the RS treatment.

\begin{figure}[h]
\centering
\includegraphics[width=8cm]{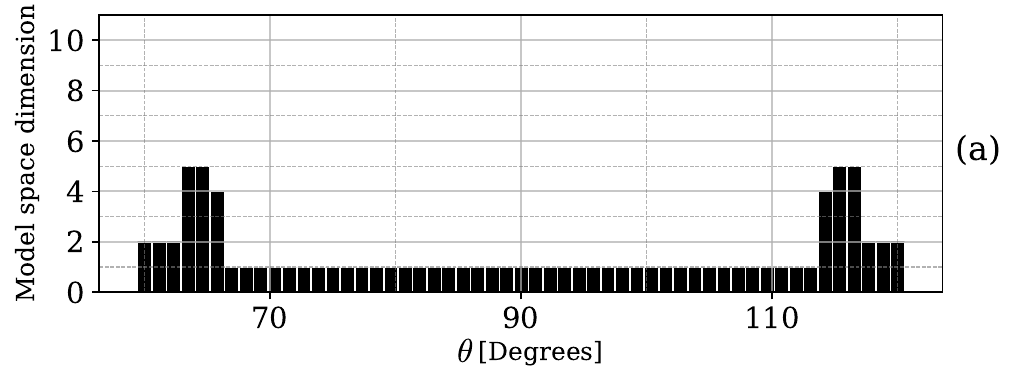}
\includegraphics[width=6.3cm]{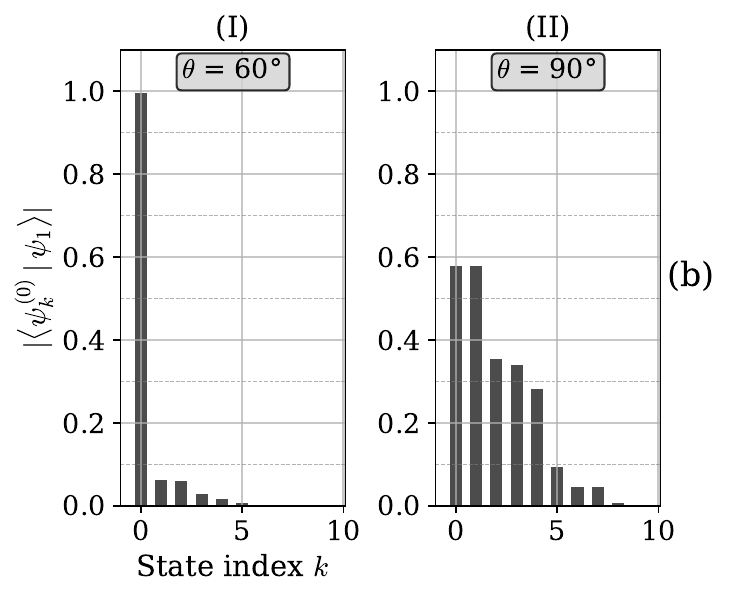}
\includegraphics[width=6.3cm]{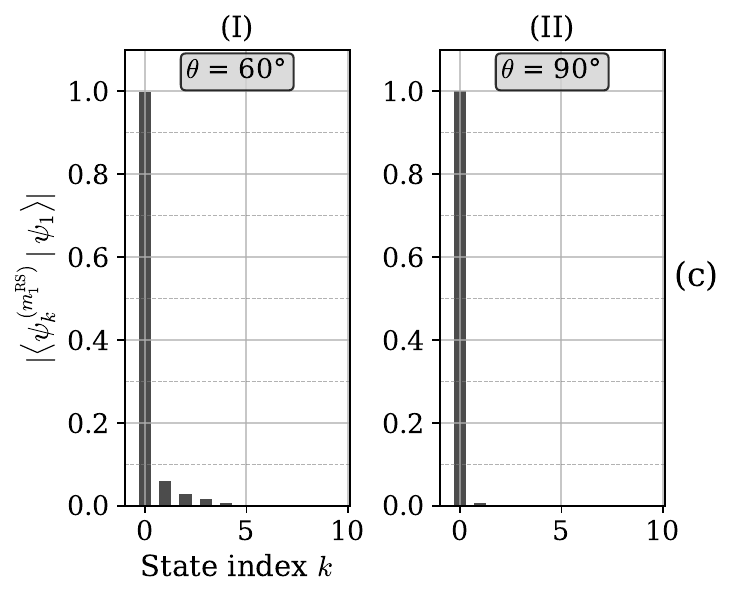}
\caption{\footnotesize Optimization of the ZO-CS for the first excited singlet state: (a) dimensions of the model spaces $P_1$ as a function of $\theta$; (b) and (c) decomposition of the FCI first excited state over the initial CSF basis and the optimized zeroth-order basis \( \mathcal{B}^{(m_1^{\rm RS})} \), respectively, shown for (I) rectangular and (II) square geometries. In the optimized basis, the projection onto the candidate state is indexed by $k=0$, while the projections onto the other states are ordered in decreasing absolute value. In the CSF basis, projections onto all states are ordered by decreasing absolute value.}
\label{fig:H4-1st}
\end{figure}

\underline{Excited singlet states:}  
For most geometries, the ZO-CS extracted from the previously optimized basis \( \mathcal{B}^{(n_0^{\rm RS})} \) already provides an optimal description, making additional RS treatments unnecessary and reducing the dimensions of the model space $P_1$ as compared to $P_0$ (see Figure~\ref{fig:H4-1st}-a).
However, in the vicinity of $\theta = 66^\circ$ (and its symmetric counterpart $\theta = 114^\circ$), 
the dimension
of the model space increases with two (quasi-)degenerated
excited singlet states. This point will be addressed later in this section. 
As for the ground state,
the decompositions of the first excited state on 
\( \mathcal{B}^{(0)} \)
shown in Figure~\ref{fig:H4-1st}-b 
extend over several components, particularly in the square geometry (see Figure~\ref{fig:H4-1st}-b(II)).
Finally, Figure~\ref{fig:H4-1st}-c stresses the compact 
expansion of the FCI excited state onto the optimized ZO-CS, favouring the
subsequent BW expansion.

Similar conclusions can be drawn for the second excited state. 
First, the model space is significantly reduced as
compared to the decomposition  onto the \( \mathcal{B}^{(0)} \) basis.
Second, the optimized ZO-CS strongly overlap with the corresponding FCI eigenstate. 
Therefore, the effectiveness of the RS procedure in constructing a representative reference state is confirmed for the second singlet excited state as well.

\subsection{BW-corrected energies}
\begin{figure}[H]
\centering
\includegraphics[width=5.cm]{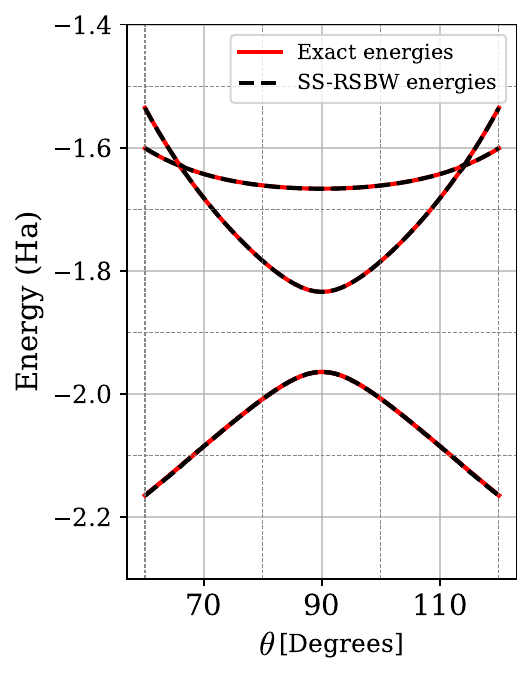}
\caption{\footnotesize SS-RSBW energies \( \{E^\text{SS-RSBW}_{0\text{–}2}\} \) 
of the ground and two lowest singlet excited states of the H$_4$ ring as a function of $\theta$. Exact FCI results \( \{E^{\mathrm{exact}}_{0\text{–}2}\} \) are shown for comparison.  The first and second excited states crossings for $\theta = 66^\circ$ (and symmetric counterpart $\theta = 114^\circ$) are accurately reproduced by the SS-RSBW procedure.}
\label{fig:Fig-H4-energies}
\end{figure}
\renewcommand{\arraystretch}{1.3} 
\begin{table}[H]  
 \begin{center}
       \begin{tabular}{c|c|c|c}
    \hline
    Energy Level & \( E_0 \) & \( E_1 \) & \( E_2 \) \\    
    \hline \hline
    \( \Delta^{\rm RS} \) (mHa) & \( 0.4 \) & \( 0.02 \) & \( 0.08 \) \\
    \hline 
    \( \Delta^\text{SS-RSBW} \) (mHa) & \( 0.04 \) & \( 0.01 \) & \( 0.01 \) \\
    \hline    
  \end{tabular}
  \caption{Root-mean-square  errors (mHa) for the three lowest singlet energy levels of H$_4$. 
The first row reports errors from RS treatments, while the second row corresponds to the SS-RSBW results. 
References are FCI values.
  }
  \label{tab:errors-H4}
 \end{center}
\end{table}
Figure~\ref{fig:Fig-H4-energies} shows the FCI energies of the ground state and the two lowest singlet excited states as a function of $\theta$, together with the SS-RSBW results (dashed line). The SS-RSBW energies were obtained from a second-order BW expansion and converged with less than five iterations (see Eq.~\ref{eq:BW_general}). Overall, the agreement is excellent (see Table~\ref{tab:errors-H4}) for both SS-RSBW and RS energies across the entire angular range considered.


For $\theta =  66^\circ$ and $\theta =114^\circ$, the two excited states
are degenerate, a feature which is reproduced with high accuracy by our approach. 
In practice, the procedure triggers a local enrichment of the model space  during the optimization of the ZO-CS for the first excited state (see Figure~\ref{fig:H4-1st}-a), thereby explicitly including the CSFs responsible for the coupling of the two states. This targeted enlargement of the model
space improves the conditioning of the BW correction and enables the method to recover two numerically degenerate levels, whereas a naïve fixed-reference perturbative strategy would likely fail. This behavior has also been analyzed in previous works on this method \cite{bindech2024combining}.

\section{Conclusion}

We have introduced and applied a perturbative framework that combines
an iterative Rayleigh–Schrödinger (RS) optimization of zeroth-order states
with second-order Brillouin–Wigner (BW) correction. 
This two-step SS-RSBW method enables the construction of compact and refined
reference wavefunctions that concentrate the  
multiconfigurational character of both ground and excited states. By targeting individual eigenstates successively, and optimizing one at a time their zeroth-order representation, the method offers a controlled and accurate alternative to traditional multi-state approaches.

Applications to the LiH and H$_4$ ring  molecules demonstrate the effectiveness of the proposed scheme in both accuracy and computational efficiency. Throughout the geometry changes, 
the optimized zeroth-order candidate states consistently exhibit dominant overlap with the exact FCI solutions. Whatever the geometries, the BW-corrected energies for the three lowest singlet states of LiH and H$_4$ show very good agreement with FCI reference values, with root-mean-square (RMS) errors consistently lower than \(10^{-4}\) Hartree.
These results underscore the ability of the RS treatment to isolate relevant part of correlation effects into low-dimensional model spaces, enabling efficient, yet highly accurate, energy predictions even in strongly correlated regimes.
While targeted molecules remain relatively small systems for which full FCI calculations are still tractable, it is to
be considered as a valuable benchmark  
for demonstrating the reliability and underlying principles of the SS-RSBW strategy. The compactness of the model spaces and the rapid convergence observed in this study suggest that the method can scale efficiently with system size and complexity. Future work will explore the extension of this approach to larger molecular systems and more challenging electronic structures, including 
multi-state potential energy surfaces relevant to photophysical and photochemical processes.

\section{Acknowledgments}
This work was supported by the ANR (PRC ANR-2023-CE07-0035 AtropoPhotoCat).
The authors wish to thank Prof. Celestino Angeli and 
Dr. Jean-Paul Malrieu for helpful discussions regarding
the transferability of the method.


\section*{References}
\bibliographystyle{iopart-num} 
\bibliography{biblio}

\end{document}